\documentclass[fleqn,usenatbib]{mnras}

\usepackage[T1]{fontenc}
\usepackage{ae,aecompl,color,ulem}
\usepackage{booktabs}   
\usepackage{multirow}

\usepackage{graphicx}	
\usepackage{amsmath}	
\usepackage{amssymb}	

\newcommand{\thetaobs}{\theta_v}

\newcommand{\afterglowpy}{\texttt{afterglowpy}}

\title[X-ray lightcurve of GW170817]{
Accurate flux calibration of GW170817: is the X-ray counterpart on the rise? 
}

\author[Troja et al.]{
E. Troja$^{1,2}$\thanks{E-mail: eleonora@umd.edu}, B. O'Connor$^{1,2,3,4}$, 
G. Ryan$^{1,2}$, L. Piro$^5$,   R. Ricci$^{6,7}$, B. Zhang$^8$,
\newauthor
T. Piran$^9$, G. Bruni$^5$, S. B. Cenko$^{2,10}$, H. van Eerten$^{11}$
\\
$^{1}$ Department of Astronomy, University of Maryland, College Park, MD 20742-4111, USA \\
$^{2}$Astrophysics Science 
Division, NASA Goddard Space Flight Center, 8800 Greenbelt Rd, Greenbelt, MD 20771, USA\\
$^{3}$ Department of Physics, The George Washington University, 725 21st Street NW, Washington, DC 20052, USA\\
$^{4}$ Astronomy, Physics, and Statistics Institute of Sciences (APSIS), The George Washington University, Washington, DC 20052, USA\\
$^{5}$INAF -- Istituto di Astrofisica e Planetologia Spaziali, via Fosso del Cavaliere 100, I-00133 Roma, Italy\\
$^{6}$ Istituto Nazionale di Ricerche Metrologiche – Torino, Strada delle Cacce 91, I-10135 Torino, Italy\\
$^{7}$ INAF -- Istituto di Radioastronomia, via Gobetti 101, I-40129 Bologna, Italy\\
$^{8}$ Department of Physics and Astronomy, University of Nevada, Las Vegas, NV 89154, USA \\
$^{9}$ Racah Institute of Physics, Edmund J. Safra Campus, Hebrew University of Jerusalem, Jerusalem 91904, Israel\\
$^{10}$ Joint Space-Science Institute, University of Maryland, College Park, Maryland 20742, USA \\
$^{11}$Department of Physics, University of Bath, Claverton Down, Bath BA2 7AY, United Kingdom\\
}
\date{Accepted XXX. Received YYY; in original form ZZZ}

\pubyear{2018}

\begin{document}

\pagerange{\pageref{firstpage}--\pageref{lastpage}}
\maketitle
\label{firstpage}

\begin{abstract}
X-ray emission from the gravitational wave transient GW170817 is well described
as non-thermal afterglow radiation produced by a structured relativistic jet 
viewed off-axis.  We show that the X-ray counterpart continues to be detected
at 3.3 years after the merger. Such long-lasting signal is not 
a prediction of the earlier jet models characterized by a narrow jet core
and a viewing angle $\approx$20~deg, and is spurring a renewed interest in the
origin of the X-ray emission. 
We present a comprehensive analysis of the X-ray dataset aimed at 
clarifying existing discrepancies in the literature, and in particular 
the presence of an X-ray rebrightening at late times.
Our analysis does not find evidence for an increase in the X-ray flux,
but confirms a growing tension between the observations and the jet model. 
Further observations at radio and X-ray wavelengths would be critical 
to break the degeneracy between models.

\end{abstract}

\begin{keywords}
stars: neutron -- gravitational waves -- gamma-ray burst  
\end{keywords}



\section{Introduction}





The ground-breaking discovery of the binary neutron star (BNS) merger GW170817 by the LIGO/VIRGO Collaboration \citep{gw170817} and the near-coincident detection, with a delay of 1.7 s, of the short duration gamma-ray burst GRB 170718A \citep{LVCGBM} heralded a new era of multi-messenger astrophysics combining gravitational waves (GW) with photons. 
GRB 170817A, at a distance of only $\sim$\,$40$ Mpc, 
is the least luminous short GRB known to date. 
It does not display the standard fading afterglow of GRBs, 
but a delayed X-ray \citep{Troja2017} and radio \citep{Hallinan2017} emission. 
Its broadband afterglow is seen to rise as $F_\nu\propto t^{0.8}$ \citep{Troja2018,Lyman2018, Margutti2018,Ruan18}, 
peak at $\sim$\,$160$ d after the merger \citep{Dobie2018,DAvanzo2018,Piro2018}, and then rapidly decay as $F_\nu\propto t^{-2.2}$ \citep{Mooley2018superluminal,Lamb2019,Troja19}.

The afterglow behavior is now commonly interpreted as emission from a structured GRB jet viewed off-axis, with viewing angle $\theta_v$\,$\approx$\,20-30\,deg 
\citep{Troja2017,Lazzati2018,Lyman2018,DAvanzo2018,
Xie2018,Margutti2018,Resmi2018, Mooley2018superluminal, Ghirlanda2018,Lamb2019,Ryan19,Troja19,Beniamini2020,Nathanail2020,troja20,MAK20}. 
The close distance of the event and its bright long-lived emission 
allowed for an unprecedented insight into the structure of GRB jets 
and novel constraints on the Hubble Constant \citep{Hotokezaka19hubblecon,Nakar2021}. 
Continued monitoring of the GW afterglow will further deepen our understanding of GRB physics into a poorly explored regime. 
Whereas the rising slope of the light curve is dictated by the initial jet structure and the viewing angle \citep{Ryan19,Takahashi2020,Takahashi2021}, its late-time evolution (postpeak) will be dictated by the spreading dynamics of the jet and its deceleration into a non-relativistic flow. Although the measured decay slope is sufficiently steep to confirm the presence of a collimated jet \citep{Troja2018}, the exact predicted slope at this stage remains sensitive to details in the modeling and to the detailed features of the actual outflow. Other factors can impact the slope as well \citep{troja20}, such as 
a change in the properties of particle-shock acceleration across the transition from relativistic to non-relativistic shocks.

Most interestingly, now that the emission from the relativistic jet is fading away, new emission components may become 
visible \citep{troja20,corsi21,Hajela21}. 
A popular model is the so called ``radio flare"  - non-thermal radiation produced by the deceleration of the fastest merger ejecta \citep{np11,Hotokezaka18}, also referred to as kilonova afterglow \citep{kathi18}. 
This new component would appear as a slowly rising radio counterpart, 
visible a few year after the merger,
although interaction between the relativistic jet and the merger ejecta 
may quench it and further delay its onset \citep{Margalit2020,Ricci21}. 
Depending on the spectral shape of the radio flare, 
its signal may also be detectable at X-ray energies \citep{kathi18,Hajela19,troja20}. 
Late-time emission from the central compact object was also discussed \citep{Murase18,Piro2018}, and could unveil the nature of the elusive merger remnant.

Any deviation from the relativistic  structured jet model  is of great interest, 
whether it belongs to the jet dynamical evolution,  
the changing nature of particle acceleration once shocks enter the trans-relativistic regime, or to the emergence of an additional components. However, its identification is complicated by the faintness of the source which, at this point in time, is only marginally detectable with  the existing instrumentation. Different statistical treatments of the low-count regime
and/or different modeling of the instrumental effects might introduce
a systematic uncertainty in the flux measurements. 
This issue seems to be particularly relevant for the X-ray fluxes reported in the literature with values differing by up to a factor of two for the same dataset.

In this work, we present a homogeneous re-analysis of the X-ray dataset aimed at
characterizing such differences and, in particular, at addressing  the onset of a new component of emission at $\approx$3 yr post-merger,
as discussed in \cite{troja20} and recently more firmly claimed by \citet{Hajela21}. 
In \S \ref{sec:obs}, we present the observations and data analysis. In \S \ref{sec: jet}, we discuss a comparison of the jet model to these latest observations, and in \S \ref{sec: conclusion} we summarize our findings. Throughout this paper, 
times are referenced to the GRB trigger. 
We adopt a standard $\Lambda$CDM cosmology \citep{Planck2018}. Unless otherwise stated, the quoted errors are at the 68\% confidence level, and upper limits are at the 3\,$\sigma$ confidence level.

\section{Observations and data analysis}
\label{sec:obs}

The target has been regularly monitored with the \textit{Chandra} X-ray Telescope
starting on August 19, 2017 ($T_0$+ 2.3~d) until January 27, 2021 ($T_0$+ 1258.7~d). 
The entire dataset, consisting of 31 observations spread over 11 epochs, 
was reprocessed using the latest release of the 
Chandra Interactive Analysis of Observations (CIAO v.~4.13; \citealt{fruscione06}) and calibration files (CALDB 4.9.4). 

We follow the same steps described in \citet{troja20}, including background filtering and astrometric alignment of each observation. 
Aperture photometry was performed in the broad 0.5- 7.0 keV energy band. 
Since the target is placed close to the optical axis, the point spread function (PSF) can be considered symmetric and source counts are extracted using a circular aperture with radius of 1.5$\arcsec$. If less than 15 counts are extracted, we use a smaller radius of 1.0$\arcsec$ in order to optimize the signal to noise ratio. 
Aperture corrections are derived through the task \texttt{arfcorr} and are typically $\lesssim$1.1. 
The background level is estimated from nearby source-free circular regions with radius $\gtrsim$15$\arcsec$.

\begin{figure}
\includegraphics[width=\columnwidth]{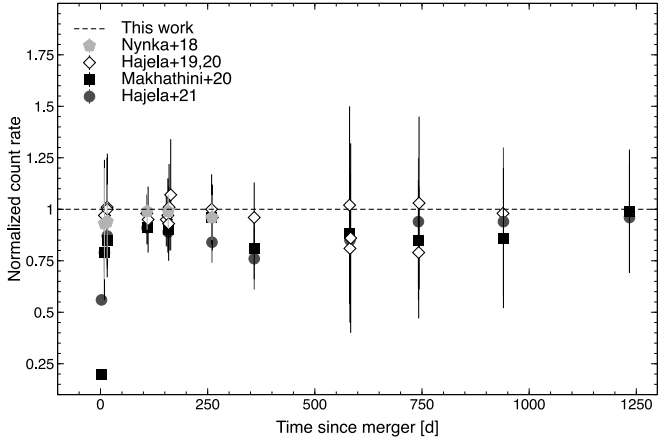}
\caption{Count-rates presented in the literature 
\citep{Nynka,Hajela19,Hajela20,MAK20,Hajela21}
normalized by the values derived in this work. 
This comparison shows an overall agreement of the 
different analyses, except for the first data point at 2.3~d.
Other differences may depend on the aperture correction and whether
the reported count-rates include it or not.}
\label{fig:rate}
\end{figure}

The final net count rate is then derived as  
$r_s = \eta\,(N - B \times A_s / A_b ) \Delta t^{-1}$, where $N$ and $B$ are the 
measured total and background counts within the extraction regions of area
$A_s$ and $A_b$, respectively; $\eta$ is the energy-dependent aperture correction, and 
$\Delta t$ the exposure time of each observation. 
The detection significance and confidence intervals on the count rates are calculated following \citet{kbn91}. 
Except for the first observation at 2.3~d (ObsID 18955), X-ray emission from the position of GW170817 
is detected at all epochs with significance $\gtrsim$3\,$\sigma$.
A comparison between our results and the values reported in the literature 
\citep{Hajela21,MAK20,Hajela20,Hajela19,Nynka} shows
an overall consistency of the derived count rates (Figure~\ref{fig:rate}).
The values of \citet{Hajela21} and \citet{MAK20} (priv. comm.) appear systematically lower
by a factor $\approx$1.1, a value consistent with the aperture correction applied in this work. 
A discrepancy worth of note is the upper limit at 2.3~d. Within our source extraction region, we measure zero counts in a 24.6 ks exposure, 
from which we derive a 3\,$\sigma$ upper limit of 2.5$\times$10$^{-4}$ cts\,s$^{-1}$, twice the value reported in other works 
\citep{Hajela21,Hajela19,Nynka}
and five times higher than the value quoted in \citet{MAK20}. 
Since we already measure the minimum number of counts,
we attribute this difference to the statistical treatment 
of upper limits. Our limit is derived using the formulation of \citet{kbn91}, and a similar value is obtained using
the approximations of \citet{Gehrels86}.
Our results are listed in Table~\ref{tab1} for each epoch, 
whereas the detailed analysis of each observation is reported in the Appendix (see Table~\ref{tab3}). 

\begin{figure}
\includegraphics[width=\columnwidth]{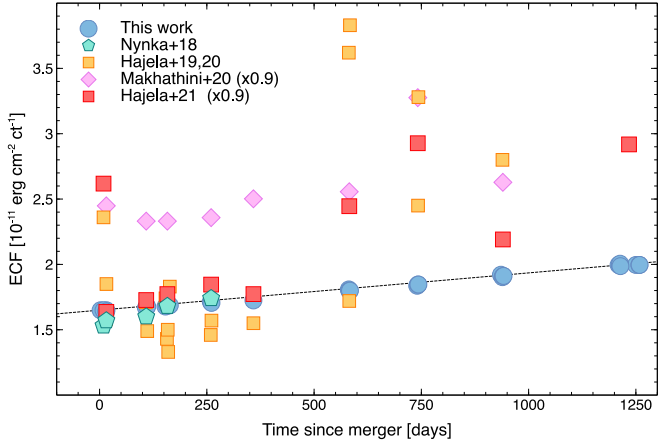}
\caption{Energy conversion factor (ECF) used to transform
count rates into fluxes in the 0.3-10 keV band. 
Our results are compared with values derived from the literature
\citep{Nynka,Hajela19,Hajela20,MAK20,Hajela21}, 
highlighting substantial differences in the reported fluxes.}
\label{fig:ecf}
\end{figure}

We then convert the observed count-rates into X-ray fluxes by folding the 
afterglow spectral shape with the instrumental response. 
A joint spectral fit of the radio, optical, and X-ray data shows a
power-law spectrum with photon index $\Gamma$=1.585$\pm$0.005 \citep{Troja19} and
negligible intrinsic absorption in addition to the Galactic value 
of 1.1$\times$10$^{21}$\,cm$^{-3}$ \citep{Willingale13}.
We therefore use this model to derive an energy conversion
factor (ECF) for each observation. 
We find that observations performed within a few days of each other
presents negligible differences in their ECF.
However, the entire observing campaign spans over three years and an appreciable
increase of the ECF is visible, from $\approx$1.7$\times10^{-11}$ in 2017 
to $\approx$2.0$\times10^{-11}$ in 2021. 
The resulting ECFs and X-ray fluxes, calculated for a constant spectral index, are listed in Table~\ref{tab1} in the ``$\Gamma$=1.585'' columns. 

Figure~\ref{fig:ecf} compares our values to the results of \citet{Hajela21},  \citet{Hajela20}, \citet{MAK20},
\citet{Hajela19} and \citet{Nynka}, who also present a comprehensive re-analysis of the X-ray afterglow data. The ECFs were derived by dividing the reported fluxes for their respective count-rates:
in the case of \citet{Nynka}, the unabsorbed X-ray fluxes were derived by rescaling their
luminosity values; in the case of \citet{MAK20}, we 
rescaled the reported flux densities at 1~keV (in $\mu$Jy) by a factor of 
1.18$\times$10$^{-11}$ calculated for a photon index of 1.57; 
in the case of \citet{Hajela21} and \citet{MAK20}, the ratio between fluxes and count-rates 
is further divided by $\approx$1.1 in order to account for PSF losses in 
a 1$\arcsec$ radius aperture. 
If an aperture correction is already applied to their reported count rates, 
the discrepancy would be larger. 

As shown in Figure \ref{fig:ecf}, we find a good agreement with the values of \citet{Nynka} and, partially, with those of \citet{Hajela21} between 15~d and 260~d (Epochs 3-6 in Table~\ref{tab1}). In other epochs the work of \citet{Hajela21} derives higher and highly variable ECFs, not consistent with our analysis. The net result is an higher average flux level at late times. 
A systematic discrepancy is also found with the values quoted in \citet[][Table~1]{MAK20}, 
which are consistently higher than our values by 40\%. 
Such large discrepancy is only found when using the flux densities at 1~keV reported in their Table~1.
By comparing the fluxes of the single \textit{Chandra} observations (our Table~\ref{tab3} and Table~2 in \citealt{MAK20}), 
we find a good agreement between the two works. 
With respect to our previous analyses \citep{Troja2017,Troja2018,Piro2018,Troja19,troja20}, we find consistent values and only note that 
the X-ray fluxes increased by 10\%  the values in \citet{troja20} 
due to the updated calibration files used in this work.

In contrast to our method, which is based on the broadband (from radio to X-rays) spectral fitting of the afterglow data, \citet{Hajela19} and \citet{Hajela21} determine the spectral shape, and hence the ECFs,
using only the X-ray data.
This method has some advantages: it is independent from the afterglow model and potentially sensitive to the source spectral evolution, but in practice it is dominated  by the large uncertainties of the low-counts regime.
Nonetheless, for the sake of comparison, we also calculate the ECFs for the case of a time-variable spectral index. 
For several observations, and in particular those at early ($<$20~d) and late ($>$1 yr) times, we do not have sufficient photons for spectral analysis and we therefore use the hardness ratio to estimate the spectral shape \citep[e.g.][]{BurstAnalyzer}. 

\begin{figure}
\includegraphics[width=\columnwidth]{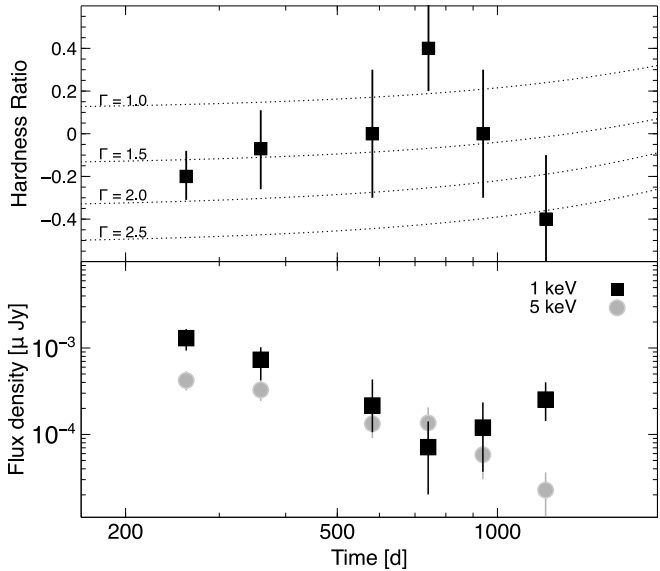}
\caption{\textit{Top panel}: temporal evolution of the hardness ration $HR$.
Dotted lines show the values expected for an absorbed power-law model with photon index $\Gamma$ between 1.0 and 2.5, and take into account the evolving instrumental response.
\textit{Bottom panel}: X-ray flux light curves at 1 keV (squares) and 5 keV (circles)
derived using a time-variable photon index inferred from X-ray observations.
The apparent rise of the soft X-ray emission (1 keV) is a result of the
hard-to-soft spectral evolution seen at late times.}
\label{fig:hr}
\end{figure}

We define the hardness ratio as $HR$=$(H - S)/(H + S)$, where H and S are the net source counts in the hard (2.0-7.0 keV) and soft (0.5-2.0 keV) energy bands, respectively.
Its late-time temporal evolution is shown in Figure~\ref{fig:hr}, updated
from \citet{troja20} using the latest observations at
$\approx$1230~d and the relevant calibration files. 
We still assume an absorbed power-law model with $N_H$ fixed to the Galactic value and variable photon index $\Gamma$. 
Following \citet{BurstAnalyzer}, we input the spectral model and response files into 
the CIAO tool \texttt{modelflux}
and create a look-up table of hardness ratios and ECFs by stepping $\Gamma$ from 0 to 3 in steps of 0.1 and recording at each step the model count-rates and fluxes in different bands, namely 0.5-2.0~keV (soft), 2.0-7.0~keV (hard), and 0.5-7.0~keV (broad). 
We then derive the observed hardness ratio following \citet{Park2006}, 
and infer the corresponding photon index and ECF from the look-up table. 
The 68\% confidence level uncertainty on the $HR$ is used to estimate the error on the ECF. 
Using data from Epoch 4 ($t$=109~d), when the afterglow is sufficiently bright for an 
independent spectral analysis, 
we verify  that the photon indices, $\Gamma$=1.6$\pm$0.2 from the $HR$ and
$\Gamma$=1.66$\pm$0.17 from the spectral fit, are in good agreement.

The resulting ECFs and X-ray fluxes, calculated for a time-variable photon index, are listed in Table~\ref{tab1} in the  ``$\Gamma$ free'' columns. 
Although this method yields a better agreement with the results of \citet{Hajela21}, it cannot reproduce the increase in flux at 1230~d.
For the range of spectral indices $\Gamma$\,$\approx$1-2 typical of an afterglow, variations in the ECFs are $\lesssim$20\%. 
Spectral variations, unless extreme, do not significantly affect the flux estimates,
but can have a noticeable impact on the derived flux densities, as shown in the
bottom panel of Figure~\ref{fig:hr}. 
For a central energy of 1~keV, the conversion factor from rate to flux density 
increases by 65\% between $\Gamma$=1 and
$\Gamma$=1.5, and more than doubles between $\Gamma$=1 and $\Gamma$=2.
By suppressing the flux density in the case of a hard spectrum 
and boosting it in the case of a soft spectrum, 
the soft-hard-soft evolution seen in the HR diagram is 
at the origin of the apparent rise of the light curve at 1~keV.
This temporal feature is not seen in either the count rate,
the integrated flux or the flux light curve at 5 keV, 
which is less sensitive to spectral variations.  
It would therefore be inaccurate to interpret it as the onset of 
a new, spectrally harder component of emission as this trend 
appears only in the case of a significant ($\Delta\Gamma$\,$\gtrsim$0.5) 
hard-to-soft evolution of the X-ray spectrum.



Finally, we investigate whether instrumental artifacts, 
such as hot columns or bad pixels, 
lie close to the source position on the detector. 
These factors may cause large variations of the ECF,
such as the one seen in Figure~\ref{fig:ecf}. 
However, a visual inspection of the exposure maps shows that 
they do not affect the observations of GW170817. 
As seen in Figure~\ref{fig:expomap}, the combined exposure maps 
for the latest observations show that the
target was observed in optimal conditions. 
We therefore do not expect large variations of the ECF
between the different observations. 

\begin{table*}
\centering
\caption{\textit{Chandra} X-ray observations of GW170817.}
\label{tab1}
\begin{tabular}{ccccccccl}
\hline
\hline
&   &   &  & \multicolumn{2}{c}{$\Gamma$=1.585} & 
\multicolumn{2}{c}{$\Gamma$ free}  & \\
\cmidrule(lr){5-6} \cmidrule(lr){7-8} 
Epoch & T-T$_0$  &  Exposure  & Count rate$^{a}$  & ECF$^{b}$ & Flux$^{c,d}$   &   ECF$^{b}$ & Flux$^{c}$    & ObsID\\
      &   (d)    &  (ks)      & [0.5-7.0 keV]     &           & [0.3-10 keV] &              & [0.3-10 keV]  &      \\
\hline
1  & 2.33                   & 24.6     &  $<$2.5                 & 1.65 &  $<$4.1                &     --                 &	  --               &18955  	 \\[0.9mm]
2  & 9.2                    & 49.4     &  2.9$^{+0.9}_{-0.7}$    & 1.65 &  4.7$^{+1.5}_{-1.2}$   & 1.92$^{+0.8}_{-0.3}$	&   5.8$^{+2}_{-1.4}$    &   19294	     \\[0.9mm]
3  & 15.7                   & 93.4     &  3.4$^{+0.7}_{-0.7}$    & 1.65 &  5.6$^{+1.2}_{-1.2}$   & 1.56$^{+0.19}_{-0.10}$ &   5.3$^{+1.2}_{-1.1}$ &   20728, 18988       \\[0.9mm]
4  & 108.7                  & 98.8     &  14.9$^{+1.2}_{-1.2}$   & 1.67 &  25$^{+2}_{-2}$        & 1.66$^{+0.11}_{-0.06}$ &    25$^{+3}_{-2}$     &   20860, 20861        \\[0.9mm]
5  & 158                    & 104.9    &  15.4$^{+1.2}_{-1.2}$   & 1.69 &  26$^{+2}_{-2}$        & 1.61$^{+0.06}_{-0.02}$ &    25$^{+2}_{-2}$     &  20936, 20938, 20937, \\[0.9mm]												  
   &                        &          &                         &      &                        &                       &                       &   20939,  20945	 \\[0.5mm]
6  & 260.0                  & 96.8     &  8.2$^{+0.9}_{-0.9}$    & 1.71 &  14.0$^{+1.7}_{-1.7}$  & 1.66$^{+0.14}_{-0.04}$ &  13.6$^{+1.9}_{-1.6}$ &   21080, 21090	  \\[0.9mm]
7  & 358.6                  & 67.2     &  5.2$^{+1.0}_{-0.9}$    & 1.73 &  9.0$^{+1.7}_{-1.5}$   & 1.78$^{+0.3}_{-0.11}$  &  9.3$^{+2}_{-1.7}$   &   21371		     \\[0.9mm]
8  & 582.0                  & 98.3     &  1.7$^{+0.4}_{-0.5}$    & 1.80 &  3.1$^{+0.7}_{-0.9}$   & 1.97$^{+0.18}_{-0.2}$  &  3.4$^{+1.0}_{-1.0}$ &   21322, 22157, 22158  \\[0.9mm]
9  & 741.7                  & 98.9     &  1.1$^{+0.3}_{-0.4}$    & 1.85 &  2.0$^{+0.5}_{-0.7}$   & 2.90$^{+1.00}_{-0.6}$  &  3.1$^{+1.6}_{-1.3}$ &   21372, 22736, 22737   \\[0.9mm]
10 & 940                    & 96.6     &  0.8$^{+0.3}_{-0.3}$    & 1.91 &  1.6$^{+0.5}_{-0.7}$   & 2.00$^{+0.6}_{-0.2}$ &  1.6$^{+0.9}_{-0.7}$ &   21323, 23183, 23184   \\[0.9mm]												
   &                        &          &                         &      &                        &                       &                      &   23185		   \\[0.5mm]
11a & 1212                  & 91.1     &  1.2$^{+0.4}_{-0.3}$    & 2.00 &  2.3$^{+0.9}_{-0.6}$   & 2.30$^{+1.2}_{-0.3}$   &  2.7$^{+1.5}_{-0.8}$ &   22677, 24887, 24888    \\[0.5mm]
    &                       &          &                         &      &                        &                       &  			&   24889		  \\[0.5mm]
11b & 1255                  & 97.9     &  0.5$^{+0.2}_{-0.2}$    & 1.99 &  1.0$^{+0.5}_{-0.5}$   & 2.30$^{+1.2}_{-0.3}$   &  1.1$^{+0.8}_{-0.6}$ &   23870, 24923, 24924    \\[0.5mm]		      
    &                       &          &                         &      &                        &                       &  			&   22677, 24887, 24888,  \\[0.5mm] 
11  & 1234                  & 189      &  0.8$^{+0.2}_{-0.2}$    & 2.00 &  1.6$^{+0.5}_{-0.5}$   & 2.30$^{+1.2}_{-0.3}$   &  1.8$^{+1.1}_{-0.6}$ &   24889, 23870, 24923,   \\[0.5mm]
    &                       &          &                         &      &                        &                       &  			         & 24924		  \\[0.5mm]	
\hline
\end{tabular}
\begin{flushleft}
     \quad \footnotesize{$^a$ Count rates are  in units of 10$^{-4}$ cts s$^{-1}$. All the values are corrected for PSF losses.} \\
     \quad \footnotesize{$^b$ ECFs are in units of 10$^{-11}$ erg cm$^{-2}$ ct$^{-1}$} \\
     \quad \footnotesize{$^c$ Fluxes in units of 10$^{-15}$ erg cm$^{-2}$ s$^{-1}$. Values are corrected for Galactic extinction.} \\
     \quad \footnotesize{$^d$ The quoted values can be converted into 
     flux densities (in units of Jy) by multiplying them by a factor of 86027
      (1 keV), or 33553 (5 keV). 
      We adopt a conversion to X-ray luminosity of (2.0$\pm$0.2)$\times$10$^{53}$\,cm$^{-2}$. }
     \\
     
\end{flushleft}
\end{table*}

\begin{figure}
\includegraphics[width=0.95\columnwidth, trim=0 0 0 0, clip]{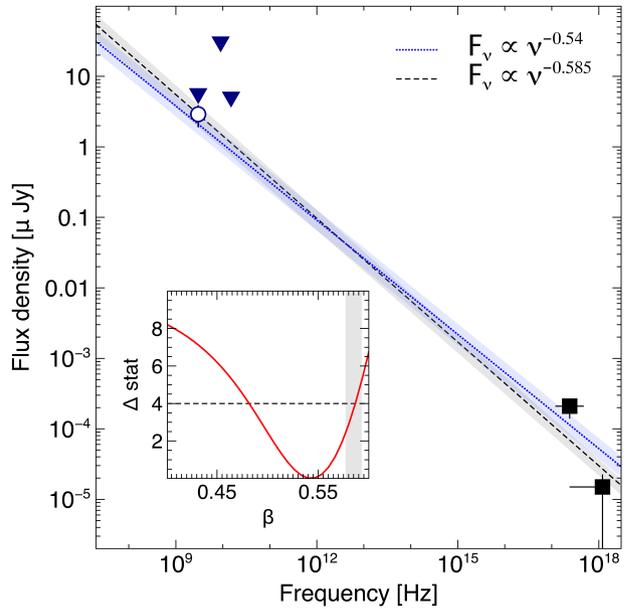}
\caption{Spectral energy distribution of the late time ($\approx$1230~d) afterglow compared with two power-law models with index 0.54 (dotted line) and 0.585 (dashed line).
The 3\,$\sigma$ radio upper limits (downward triangles) at 3 GHz, 9GHz and 15 GHz and the X-ray fluxes are derived from our analysis. The radio flux (open circle), corresponding to a marginal detection at 3 GHz, is from \citet{corsi21}.
The contour plot for the spectral index is shown in the inset. 
The afterglow value of 0.585$\pm$0.005 is marked by the vertical bar.} 
\label{fig:vla}
\end{figure}

\begin{figure*}
\includegraphics[width=2\columnwidth]{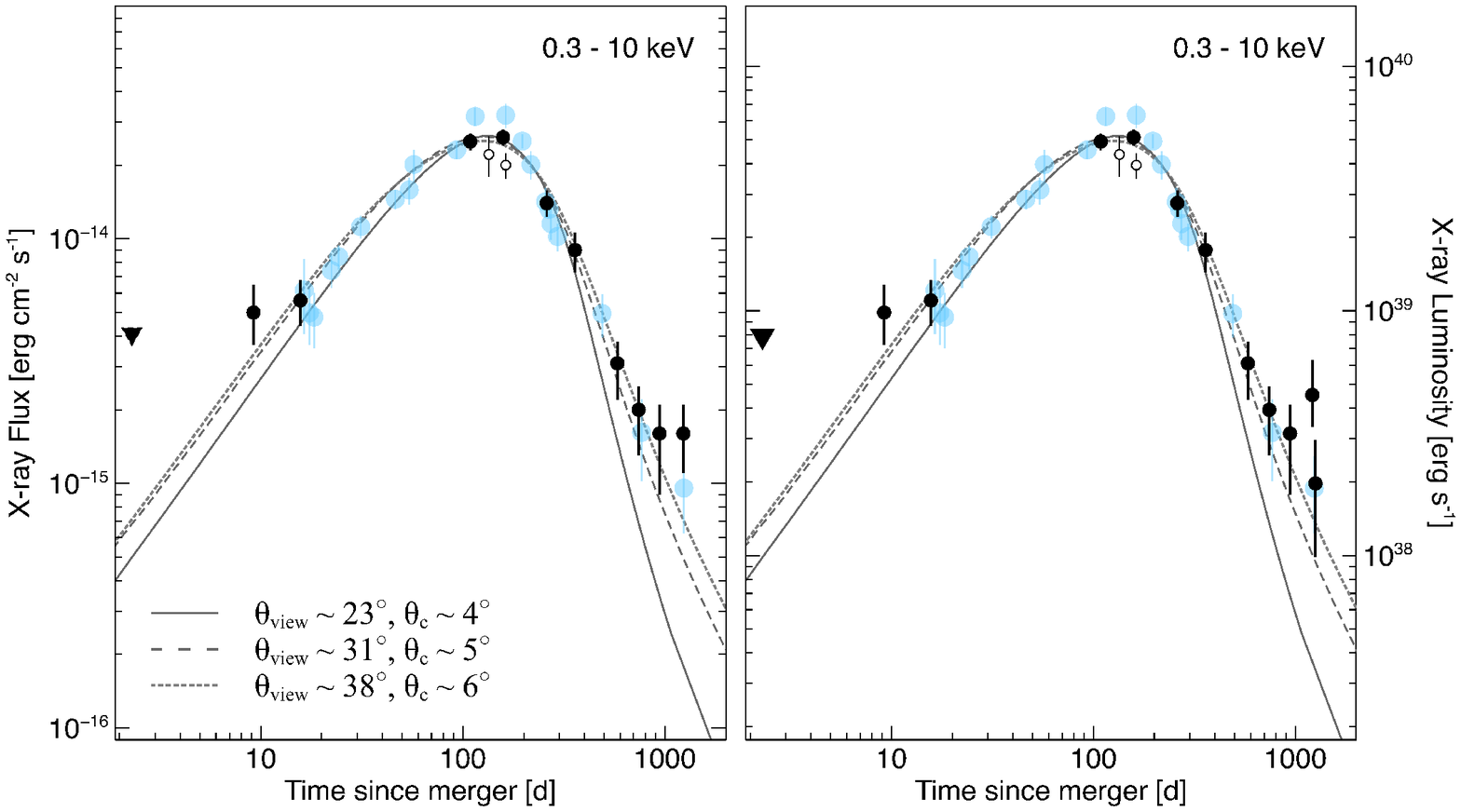}
\caption{X-ray (black circles: \textit{Chandra}; open circles: XMM-Newton) light curves compared with the jet model of 
\citet{Ryan19} (solid line), \citet{troja20} (dashed line), and this work (dotted line). Radio data (blue; \citealt{MAK20,corsi21}) at 3 GHz were rescaled 
using a spectral slope of 0.585. At late times a deviation
from the jet model is visible. By rebinning the last two \textit{Chandra} observations
(left panel), the X-ray emission seems to flatten. 
This effect is mostly driven by the detection of 
soft ($<$2 keV) X-ray emission at 1211~d, visible 
in the unbinned light curve (right panel).
}
\label{fig:jet}
\end{figure*}

\subsection{Constraints from radio observations}

We monitored the target using the Australian Telescope Compact Array (ATCA; 
project C3240, PI: L.~Piro) between November 2020 and April 2021. 
Our observations span the frequency range 2.1-9.0~GHz and are reported in Table \ref{tab4}. 
The radio counterpart is not detected and we place a 3\,$\sigma$ upper limit of 
$\lesssim$31$~\mu$Jy at 9 GHz. 

Radio observations were also carried out with the Jansky Very Large
Array (VLA) between September 2020 and February 2021, as reported in 
\citet{corsi21}. 
No signal is detected by combining $\approx$30 hrs of imaging at 3~GHz. 
By performing forced photometry at the GRB position, 
\citet{corsi21} reports a flux of 2.9$\pm$1.0 $\mu$Jy. 
We independently analyzed the public available dataset, carried out under programs SL0449 and SM0329 (PI: Margutti), totalling 12 hrs of observing time in S-band (of which $\approx$ 10 hrs on-source) and 4 hrs in Ku-band
(of which $\approx$2.3 hrs on-source). The VLA visibility data were downloaded from the NRAO online archive and calibrated with the CASA VLA pipeline v1.3.2. The splitted calibrated measurement sets from the three A-array S-band datasets (MJD 59198, 59210 and 59247) of GW170817 were merged via the CASA task {\it concat} and imaged interactively using the CASA task {\it tclean} with robustness parameter set to 0.5. Our results are listed in Table~\ref{tab4}.
The restored image is characterized by an rms of $\approx$1.9 $\mu$Jy (3 GHz) measured via the CASA task {\it imstat} in a region of the cleaned map away from sources. A similar value of $\approx$1.7 $\mu$Jy is measured in the Ku-band (15 GHz). At the position of GW170817, any visible signal is consistent with the noise level. 
At the transient position 
we find a peak force-fitted flux density of 3.1$\mu$Jy/beam. 
Our analysis is in agreement with the weak radio flux inferred by \citet{corsi21} and shows no evidence of a rebrightening in this band.

We use \texttt{XSPEC} \citep{Arnaud96} to perform a joint fit of the latest X-ray and radio data (Figure~\ref{fig:vla}). Our upper limits constrain the power-law spectral index $\beta$=$\Gamma$-1 to $\lesssim$1.6. 
The tentative radio detection of \citet{corsi21} yields $\beta$=0.54$^{+0.02}_{-0.03}$, slightly harder but consistent (within the 95\% confidence level) with the value of 0.585 derived from  afterglow spectroscopy at earlier times \citep{Troja19}. 
Using this best fit model, the X-ray flux in the 0.3-10~keV band is 1.8$^{+0.5}_{-0.6}$\,$\times$10$^{-15}$\,erg\,cm$^{-2}$\,s$^{-1}$, 
fully consistent with the value estimated in Table~\ref{tab1} (Epoch 11)
and 30-40\% lower than the flux quoted in \citet{Hajela21}. 
The low signal-to-noise of the radio and X-ray data 
does not allow us to place any strong spectral constraint. 
The slightly harder radio-to-X-ray index as well as the 
softer X-ray spectrum seen in the $HR$ diagram are both 
features of marginal statistical significance ($\approx$2\,$\sigma$ and $\lesssim$1\,$\sigma$ respectively). 

If the joint radio/X-ray analysis is performed in flux space, 
discrepancies in the flux calibration 
might  explain the different results reported in \citet{Hajela21} 
as well as the harder spectral index derived in \citet{MAK20}. 
Our fit to the X-ray data is performed in count space 
and does not depend on the flux calibration given in Table~\ref{tab1},
but is consistent with it.

\section{Comparison to the jet model}
\label{sec: jet}

Figure~\ref{fig:jet} compares the updated dataset with the jet
model presented in \citet{troja20}, who fit the dataset of the first 10 epochs
($\lesssim 940$ d) with a Gaussian structured jet. 
We have used the MCMC samples from these fits to construct posterior distributions of the model flux at 940 (Epoch 10), 1212 (Epoch 11a), 1255 (Epoch 11b), and 1234 (Epoch 11) days after the burst. 
We convert the flux predictions into counts by using the ECF (Table~\ref{tab1}, col. 5) and 
a background level of $\approx$7.8$\times$10$^{-6}$\,cts\,s$^{-1}$ within the aperture. 
The posterior predicted number of counts for each observation are 5, 3, 3, and 6 respectively. 
The corresponding observed photon counts are 8, 10, 5, and 15.  
Assuming Poissonian statistics, for each epoch we compute the probability of observing a count at least as high as the true observation, marginalized over the posterior distribution to account for uncertainty in the fit.  The 1208 day (Epoch 11a) observation displays the most significant deviation at $\approx$3$\sigma$ (Gaussian-equivalent; statistical only).  
Combining Epochs 11a and 11b into Epoch 11 at 1232 days still results in a $\approx$3\,$\sigma$ excess over the \citet{troja20} model fit. 
Epochs 11b (at 1255 days) and 10 (at 940 days) show more modest excesses of $\approx$1.2$\sigma$ each.  
These are all over-estimates of the excess over this particular jet-only model, as they do not take into account uncertainties in the calibration or modeling.

We have also performed an updated jet model fit 
including the new observations at $T > 1200$ days. The jet model is identical to that in \citet{troja20}, a Gaussian structured jet computed with \afterglowpy{} \texttt{v0.6.5} \citep{Ryan19}. With the new observations included in the fit the significance of the late time excess is reduced, as expected, at the cost of increasing tension with VLBI observations.  
Our values are lower than the significance reported by \citet{Hajela21},
showing that systematic uncertainties in the modeling of the afterglow evolution as well as in the estimates of the X-ray flux need to be taken into account. 
As shown in Figure~\ref{fig:ecf}, the higher ECF values used by \citet{Hajela21} at late times lead to higher fluxes as well as a rising temporal trend, which is not observed in count space:
in both epoch 10 (940~d) and 11 (1230~d), the source is detected at a level of 
0.8$\times$10$^{-4}$\,cts\,s$^{-1}$.

The new data confirm the trend observed in \citet{troja20}, 
a structured jet model can explain the observed X-ray emission if viewed at a larger angle than previously estimated.  The relative excess of the late-time X-ray observations 
can be accounted for by a wider jet, which has a larger total energy.   Since the afterglow's early rise at $T$\,$<$\,160~d fixes the ratio of the viewing angle to the jet opening angle \citep{Ryan19,Nakar2021}, the wider jet must be viewed proportionally further off-axis.  The new fit estimates the viewing angle $\thetaobs = 38^\circ\pm 4^\circ$, larger than the $31^\circ \pm 5^\circ$ reported in \citet{troja20} with 1000 days of data and the $23^\circ \pm 6^\circ$ reported in \citet{Troja19} and \citet{Ryan19} with 1 year of data. 
As a consequence of the larger viewing angle, the associated superluminal apparent velocity shifts from $\beta_{app}$=2.2$^{+0.5}_{-0.4}$ to
$\beta_{app} = 2.0^{+0.3}_{-0.2}$, increasing further the tension with the value of  $\beta$=4.0$\pm$0.5 determined by  the VLBI centroid motion, 
from $2.8\,\sigma$ \citep{troja20} up to $3.5\,\sigma$ when marginalized over the fit.
As noted in \citet{troja20}, the addition of an extra-component with luminosity 
$L_X$\,$\approx$\,$2 \times 10^{38}$ erg\,s$^{-1}$ would resolve this tension. 
With the additional component making up the late-time emission, the underlying jet is allowed to be narrower and nearer the line of sight, with an opening angle of $\theta_c = 4^\circ \pm 1^\circ$ and viewing angle $\thetaobs = 26^\circ \pm 6^\circ$. This alignment produces an apparent velocity of $\beta_{app} = 3.1^{+0.9}_{-0.6}$, in agreement with the measurement of \citet{Mooley2018superluminal}.

Although our analysis confirms that the X-ray and radio emission deviate from 
early predictions of the jet model with $\thetaobs$\,$\approx$20$^\circ$, 
the interpretation of this late-time behavior remains ambiguous. 
The flattening of the X-ray light curve, seen in the right panel of Figure~\ref{fig:jet}, is suggestive of an additional component taking over the fading GRB afterglow. 
Although tantalizing, the observed trend is driven mostly by a single data point at 1211~d,
deviating $\lesssim$3~$\sigma$ from the afterglow predictions (left panel of Figure~\ref{fig:jet}),
and a continued fading of the X-ray and radio counterpart remains consistent with the observations.

In \citet{troja20}, we already discussed in detail the possible origins of the late-time X-ray emission and made predictions about its future evolution. 
Here we briefly review them in light of the new observations. 
A deviation from the simpler jet model could be caused by a change in the jet dynamics. 
In the current phase of evolution the jet is trans-relativistic and undergoing lateral spreading. As noted by \citet{troja20}, a mere factor of four in density change beyond a parsec would lead to a factor of two increase in flux, both in the relativistic and non-relativistic regimes. During spreading, models in the relativistic limit show the flux to be effectively insensitive to density ($F_\nu \propto n^{(3-p)/12}$, \citealt{Granot18, Hajela21}), implying a far more drastic gradient to reproduce the observed flux. On the other hand, this would in turn hasten the onset of the non-relativistic stage where
$F_\nu \propto n^{0.4}$ \citep{Leventis2012}. 

Evolution in the properties of the non-thermal electrons, for instance a decrease in the electron index $p$ towards the expected non-relativistic value of 2 \citep{Bell78, Blandford78} as the jet decelerates, could in principle increase the X-ray flux above the fixed-$p$ predictions of our current models.  However, the full behaviour of the electron population in such an evolving-$p$ scenario is unknown, so no robust predictions, even whether the X-ray flux would increase or decrease, are possible at this time.

An exciting possibility would be emission from the counter-jet - the one pointing out in the opposite direction. As shown in \citet{troja20}, this does not arise with natural parameters - that is with a jet and circumburst medium with similar properties to those observed in our direction. 
In order to be visible the counter jet must slow down faster than the jet pointing towards us, either because of a significant density gradient in the opposite direction or possibly 
a lower counter-jet energy (Nakar, priv. comm.).

The most natural scenario is the onset of the late time flare arising from the interaction of the merger ejecta with the surrounding matter \citep{np11}.  
This signal would rise on a time scale comparable to the observation time scale with a rising slope that depends on the  velocity profile of the ejecta $m(v)$ \citep{np11,Piran2013,Hotokezaka2015}. 
To  avoid quenching by the jet blast wave \citep{Margalit2020,Ricci21}, 
this model would require a small amount $\sim 10^{-6}$ M$_\odot$
of fast moving $\sim 0.8$c material to be ejected along the polar axis. 
This high velocity could also explain the relatively early appearance of this signal. 
The spectrum of this new component should be more or less similar to the jet afterglow spectrum as the physics of the shocks that produce both is similar. Still some minor spectral changes are reasonable, but in particular we should expect a comparable or even higher increase in the radio band which, at present, is not observed.

An alternative possibility is emission from the central compact object. 
The scenario of a long-lived NS was already discussed in \citet{troja20}, \citet{Piro2018}, and references therein. This model predicts a flattening of the late-time emission as a possible signature of the inner engine.  If this signal is powered by the NS spindown energy, the observed timescales imply a poloidal field $B$\,$\approx$10$^{11}$-10$^{12}$G,
consistent with the limits set by the broadband observations \citep{ai20}. 

There are two possibilities for such a scenario: one is that the external shock is continuously energized by the pulsar wind \citep{Zhang2001}, which also predicts an achromatic signature between X-ray and radio bands. Alternatively, X-ray emission can be produced by the internal dissipation of the pulsar wind, which would not predict a simultaneous re-brightening of the radio flux \citep{Troja07}. If such a chromatic behavior is observed, it would lend strong support to the existence of a late central engine.  
Short timescale X-ray variability would be another key signature for this model. 

As the last X-ray detection appears rather soft in spectrum (Figure~\ref{fig:hr}) and 
its luminosity is comparable to the Eddington luminosity of a solar mass object, 
another possibility would be X-ray emission from fallback matter 
\citep{Rosswog07, RossiBegelman}.
In the latter case, the expected spectrum would be approximately thermal, peaking in the soft X-rays ($\lesssim$2.0~keV) and with a negligible radio signal. One can expect this component to decrease slowly  on a time scale dictated by accretion processes or by the fallback rate.
In this model, the central compact object can be either a NS or a solar-mass black hole.

Given the faintness of the source, it could be difficult to discern between different models 
unless the emission flattens at the current level, as envisioned in \citet{Piro2018}, or
starts to rise as in the ``radio flare'' scenario \citep{np11}. 

\section{Conclusions}
\label{sec: conclusion}

We present a comprehensive analysis of the X-ray emission from GW170817 and
find that the latest observation deviate from the simple jet model,
confirming the trend already noted in \citet{troja20} and more recently discussed
in \citet{corsi21} and \citet{Hajela21}. 
This is a robust trend, which does not depend on a single observation, 
but has been consistently observed at X-ray and, to a less extent,  
radio energies for several months. 

If interpreted as arising from the same jet that produces the afterglow so far, the 
recent data increases the tension (from 2.8\,$\sigma$ to 3.5\,$\sigma$) 
between the observed temporal profile,
which continues to favor large viewing angles $\thetaobs \gtrsim 30^{\circ}$, and
the constraints placed by the VLBI centroid motion, which instead points to $\thetaobs \lesssim 20^{\circ}$. 

Alternatively, the late-time data may indicate a new component of emission, arising
from the central compact object or from the long predicted flare \citep{np11} expected from the interaction of the ejecta with the surrounding matter. This interpretation would require some fast $\sim$\,0.8$c$ moving matter that probably arose from the dynamical ejecta.

However, we also highlight how systematic uncertainties in the 
calibration and modeling of the data may affect the conclusions. 
In particular, we do not find evidence of a rising X-ray emission in either count or flux space.  Similarly, we do not find any statistically significant spectral change.
The behavior of the late-time afterglow remains open to multiple interpretations, 
and continued monitoring at radio and X-ray wavelengths is key to 
identify the origin of such long-lasting emission from GW170817. 


\section*{Acknowledgements}

ET and BO were supported in part by the National Aeronautics and Space Administration (NASA) 
through grants NNX16AB66G, NNX17AB18G, and 80NSSC20K0389. 
LP and HJvE acknowledge support from the European Union’s Horizon 2020 Programme 
under the AHEAD2020 project (grant agreement n. 871158). 
LP was supported in part from MIUR, PRIN 2017 (grant 20179ZF5KS).
T.P. was supported by an advanced ERC grant TReX. 

\section*{Data Availability}
The data underlying this article will be shared on reasonable request to the corresponding author.

\bibliographystyle{aa}

\begin{thebibliography}{}

\bibitem[\protect\citeauthoryear{Abbott, et al.}{2017a}]{gw170817} Abbott B.~P., et al., 2017a, PhRvL, 119, 161101

\bibitem[\protect\citeauthoryear{{Abbott} et~al.,}{{Abbott}
  et~al.}{2017b}]{LVCGBM}
{Abbott} B.~P.,  et~al., 2017b, ApJL, {848, L13}

\bibitem[\protect\citeauthoryear{Ai, Gao, \& Zhang}{2020}]{ai20} Ai S., Gao H., Zhang B., 2020, ApJ, 893, 146. doi:10.3847/1538-4357/ab80bd


\bibitem[\protect\citeauthoryear{Arnaud}{1996}]{Arnaud96} Arnaud K.~A., 1996, ASPC,  17, ASPC..101


\bibitem[\protect\citeauthoryear{Balasubramanian et al.}{2021}]{corsi21} Balasubramanian A., Corsi A., Mooley K.~P., Brightman M., Hallinan G., Hotokezaka K., Kaplan D.~L., et al., 2021, arXiv, arXiv:2103.04821

\bibitem[\protect\citeauthoryear{Bell}{1978}]{Bell78} Bell, A. R., 1978, MNRAS, 182, 147

\bibitem[\protect\citeauthoryear{Beniamini, Granot, \& Gill}{2020}]{Beniamini2020} Beniamini P., Granot J., Gill R., 2020, MNRAS, 493, 3521. doi:10.1093/mnras/staa538

\bibitem[\protect\citeauthoryear{Blandford \& Ostriker}{1978}]{Blandford78} Blandford, R. P., Ostriker, J. D., 1978, APJ, 221, L29


  



\bibitem[\protect\citeauthoryear{{D'Avanzo} et~al.,}{{D'Avanzo}
  et~al.}{2018}]{DAvanzo2018}
{D'Avanzo} P.,  et~al., 2018, A\&A, {613, L1}

\bibitem[\protect\citeauthoryear{Dobie et al.}{2018}]{Dobie2018} Dobie D., Kaplan D.~L., Murphy T., Lenc E., Mooley K.~P., Lynch C., Corsi A., et al., 2018, ApJL, 858, L15. 

\bibitem[\protect\citeauthoryear{Evans et al.}{2010}]{BurstAnalyzer} Evans P.~A., Willingale R., Osborne J.~P., O'Brien P.~T., Page K.~L., Markwardt C.~B., Barthelmy S.~D., et al., 2010, A\&A, 519, A102. 




\bibitem[\protect\citeauthoryear{Fruscione et al.}{2006}]{fruscione06} Fruscione A., McDowell J.~C., Allen G.~E., Brickhouse N.~S., Burke D.~J., Davis J.~E., Durham N., et al., 2006, SPIE, 6270, 62701V. doi:10.1117/12.671760


\bibitem[\protect\citeauthoryear{Gehrels}{1986}]{Gehrels86} Gehrels N., 1986, ApJ, 303, 336. doi:10.1086/164079


\bibitem[\protect\citeauthoryear{Ghirlanda, et al.}{2019}]{Ghirlanda2018} Ghirlanda G., et al., 2019, Sci, 363, 968





\bibitem[Gottlieb et al.(2018)]{Gottlieb2018} Gottlieb, O., Nakar, E., Piran, T., et al.\ 2018, \mnras, 479, 588. doi:10.1093/mnras/sty1462




\bibitem[\protect\citeauthoryear{Granot, et al.}{2018}]{Granot18} {Granot}, J., {Gill}, R., {Guetta}, D., and {De Colle}, F., 2018, MNRAS, 481, 1597


\bibitem[\protect\citeauthoryear{Hajela, et al.}{2019}]{Hajela19} Hajela A., et al., 2019, ApJL, 886, L17

\bibitem[\protect\citeauthoryear{Hajela, et al.}{2020}]{Hajela20} Hajela A., et al., 2020, RNAAS, 4, 68

\bibitem[\protect\citeauthoryear{Hajela et al.}{2021}]{Hajela21} Hajela A., Margutti R., Bright J.~S., Alexander K.~D., Metzger B.~D., Nedora V., Kathirgamaraju A., et al., 2021, arXiv, arXiv:2104.02070

\bibitem[\protect\citeauthoryear{{Hallinan, Corsi,} et~al.,}{{Hallinan}
  et~al.}{2017}]{Hallinan2017} {Hallinan} G.,  et~al., 2017, Science, {358, 1579}

\bibitem[Hotokezaka \& Piran(2015)]{Hotokezaka2015} Hotokezaka, K. \& Piran, T.\ 2015, \mnras, 450, 1430. doi:10.1093/mnras/stv620
  
\bibitem[\protect\citeauthoryear{Hotokezaka, et al.}{2018}]{Hotokezaka18} Hotokezaka K., Kiuchi K., Shibata M., Nakar E., Piran T., 2018, ApJ, 867, 95

\bibitem[\protect\citeauthoryear{Hotokezaka et al.}{2019}]{Hotokezaka19hubblecon} Hotokezaka K., Nakar E., Gottlieb O., Nissanke S., Masuda K., Hallinan G., Mooley K.~P., et al., 2019, NatAs, 3, 940. doi:10.1038/s41550-019-0820-1


\bibitem[\protect\citeauthoryear{{Kathirgamaraju}, {Barniol Duran}  \&
  {Giannios}}{{Kathirgamaraju} et~al.}{2018}]{kathi18}
{Kathirgamaraju} A.,  {Barniol Duran} R.,   {Giannios} D.,  2018, MNRAS, {473, L121}

\bibitem[Kasliwal et al.(2017)]{Kasliwal2017} Kasliwal, M.~M., Nakar, E., Singer, L.~P., et al.\ 2017, Science, 358, 1559. doi:10.1126/science.aap9455




\bibitem[\protect\citeauthoryear{Kraft, Burrows, \& Nousek}{1991}]{kbn91} Kraft R.~P., Burrows D.~N., Nousek J.~A., 1991, ApJ, 374, 344. doi:10.1086/170124



\bibitem[\protect\citeauthoryear{Lamb, et al.}{2019}]{Lamb2019} Lamb G.~P., et al., 2019, ApJL, 870, L15


\bibitem[\protect\citeauthoryear{{Lazzati}, {Perna}, {Morsony}, {Lopez-Camara},
  {Cantiello}, {Ciolfi}, {Giacomazzo}  \& {Workman}}{{Lazzati}
  et~al.}{2018}]{Lazzati2018}
{Lazzati} D.,  {Perna} R.,  {Morsony} B.~J.,  {Lopez-Camara} D.,  {Cantiello}
  M.,  {Ciolfi} R.,  {Giacomazzo} B.,   {Workman} J.~C.,  2018,   
  [Physical Review Letters], {120, 241103}
  
\bibitem[\protect\citeauthoryear{Leventis et al.}{2012}]{Leventis2012} Leventis K., van Eerten H.~J., Meliani Z., Wijers R.~A.~M.~J., 2012, MNRAS, 427, 1329. doi:10.1111/j.1365-2966.2012.21994.x

\bibitem[\protect\citeauthoryear{{Lyman} et~al.,}{{Lyman}
  et~al.}{2018}]{Lyman2018}
{Lyman} J.~D.,  et~al., 2018, Nature Astronomy

\bibitem[\protect\citeauthoryear{Makhathini et al.}{2020}]{MAK20} Makhathini S., Mooley K.~P., Brightman M., Hotokezaka K., Nayana A., Intema H.~T., Dobie D., et al., 2020, arXiv, arXiv:2006.02382

\bibitem[Margalit \& Piran(2020)]{Margalit2020} Margalit, B. \& Piran, T.\ 2020, \mnras, 495, 4981. doi:10.1093/mnras/staa1486

\bibitem[\protect\citeauthoryear{{Margutti} et~al.,}{{Margutti}
  et~al.}{2018}]{Margutti2018}
{Margutti} R.,  et~al., 2018, ApJL, {856, L18}

\bibitem[\protect\citeauthoryear{{Mooley} et~al.,}{{Mooley}
  et~al.}{2018}]{Mooley2018superluminal}
{Mooley} K.~P.,  et~al., 2018b, Nature, {561, 355}

\bibitem[\protect\citeauthoryear{Murase et al.}{2018}]{Murase18} Murase K., Toomey M.~W., Fang K., Oikonomou F., Kimura S.~S., Hotokezaka K., Kashiyama K., et al., 2018, ApJ, 854, 60. doi:10.3847/1538-4357/aaa48a

\bibitem[\protect\citeauthoryear{Nakar \& Piran}{2011}]{np11}
Nakar E. \& Piran T., 2011, Nature, 478, 82

\bibitem[Nakar \& Piran (2021)]{Nakar2021} Nakar, E. \& Piran, T.\ 2021, \apj, 909, 114. doi:10.3847/1538-4357/abd6cd

\bibitem[Nakar (2021)]{Nakar2021a} Nakar, E., private communication, \ 2021

\bibitem[\protect\citeauthoryear{Nathanail et al.}{2020}]{Nathanail2020} Nathanail A., Gill R., Porth O., Fromm C.~M., Rezzolla L., 2020, MNRAS, 495, 3780. doi:10.1093/mnras/staa1454

\bibitem[\protect\citeauthoryear{Nynka et al.}{2018}]{Nynka} Nynka M., Ruan J.~J., Haggard D., Evans P.~A., 2018, ApJL, 862, L19. doi:10.3847/2041-8213/aad32d

\bibitem[\protect\citeauthoryear{Park, et al.}{2006}]{Park2006} Park T., Kashyap V.~L., Siemiginowska A., van Dyk D.~A., Zezas A., Heinke C., Wargelin B.~J., 2006, ApJ, 652, 610

\bibitem[Piran et al.(2013)]{Piran2013} Piran, T., Nakar, E., \& Rosswog, S.\ 2013, \mnras, 430, 2121. doi:10.1093/mnras/stt037

\bibitem[\protect\citeauthoryear{Piro, et al.}{2019}]{Piro2018} Piro L., et al., 2019, MNRAS, 483, 1912

\bibitem[\protect\citeauthoryear{{Planck Collaboration} et~al.,}{{Planck
  Collaboration} et~al.}{2018}]{Planck2018}
{Planck Collaboration} et~al., 2018, preprint,  {arXiv}
  {1807.06209}

\bibitem[\protect\citeauthoryear{Resmi, et al.}{2018}]{Resmi2018} Resmi L., et al., 2018, ApJ, 867, 57

\bibitem[\protect\citeauthoryear{Ricci et al.}{2021}]{Ricci21} Ricci R., Troja E., Bruni G., Matsumoto T., Piro L., O'Connor B., Piran T., et al., 2021, MNRAS, 500, 1708. doi:10.1093/mnras/staa3241

\bibitem[\protect\citeauthoryear{Rossi \& Begelman}{2009}]{RossiBegelman} Rossi E.~M., Begelman M.~C., 2009, MNRAS, 392, 1451. doi:10.1111/j.1365-2966.2008.14139.x

\bibitem[\protect\citeauthoryear{Rosswog}{2007}]{Rosswog07} Rosswog S., 2007, MNRAS, 376, L48. doi:10.1111/j.1745-3933.2007.00284.x

\bibitem[\protect\citeauthoryear{Ruan, et al.}{2018}]{Ruan18} Ruan J.~J., Nynka M., Haggard D., Kalogera V., Evans P., 2018, ApJL, 853, L4

\bibitem[\protect\citeauthoryear{Ryan et al.}{2020}]{Ryan19} Ryan G., van Eerten H., Piro L., Troja E., 2020, ApJ, 896, 166. doi:10.3847/1538-4357/ab93cf


\bibitem[Takahashi \& Ioka(2020)]{Takahashi2020} Takahashi, K. \& Ioka, K.\ 2020, \mnras, 497, 1217. doi:10.1093/mnras/staa1984

\bibitem[Takahashi \& Ioka(2021)]{Takahashi2021} Takahashi, K. \& Ioka, K.\ 2021, \mnras, 501, 5746. doi:10.1093/mnras/stab032


\bibitem[\protect\citeauthoryear{Troja et al.}{2007}]{Troja07} Troja E., Cusumano G., O'Brien P.~T., Zhang B., Sbarufatti B., Mangano V., Willingale R., et al., 2007, ApJ, 665, 599. doi:10.1086/519450

\bibitem[\protect\citeauthoryear{{Troja} et~al.,}{{Troja}
  et~al.}{2017}]{Troja2017}
{Troja} E.,  et~al., 2017, Nature, {551, 71}

\bibitem[\protect\citeauthoryear{{Troja} et~al.,}{{Troja}
  et~al.}{2018}]{Troja2018}
{Troja} E.,  et~al., 2018a, MNRAS, {478, L18}

\bibitem[\protect\citeauthoryear{{Troja} et~al.,}{{Troja}
  et~al.}{2019}]{Troja19} Troja E., et al., 2019, MNRAS, 489, 1919


\bibitem[\protect\citeauthoryear{{Troja}~et al.}{{Troja}
  et~al.}{2020}]{troja20} Troja E., van Eerten H., Zhang B., Ryan G., Piro L., Ricci R., O'Connor B., et al., 2020, MNRAS, 498, 5643. doi:10.1093/mnras/staa2626



\bibitem[\protect\citeauthoryear{Willingale, et al.}{2013}]{Willingale13} Willingale R., Starling R.~L.~C., Beardmore A.~P., Tanvir N.~R., O'Brien P.~T., 2013, MNRAS, 431, 394




\bibitem[\protect\citeauthoryear{{Xie}, {Zrake}  \& {MacFadyen}}{{Xie}
  et~al.}{2018}]{Xie2018}
{Xie} X.,  {Zrake} J.,   {MacFadyen} A.,  2018, ApJ, {863, 58}



\bibitem[\protect\citeauthoryear{Zhang \& M{\'e}sz{\'a}ros}{2001}]{Zhang2001} Zhang B., M{\'e}sz{\'a}ros P., 2001, ApJL, 552, L35. doi:10.1086/320255

\end{thebibliography}

\appendix
\section{Supplementary Material}

\begin{figure*}
\begin{center}
\includegraphics[width=1.5\columnwidth, trim=0 0 0 0, clip]{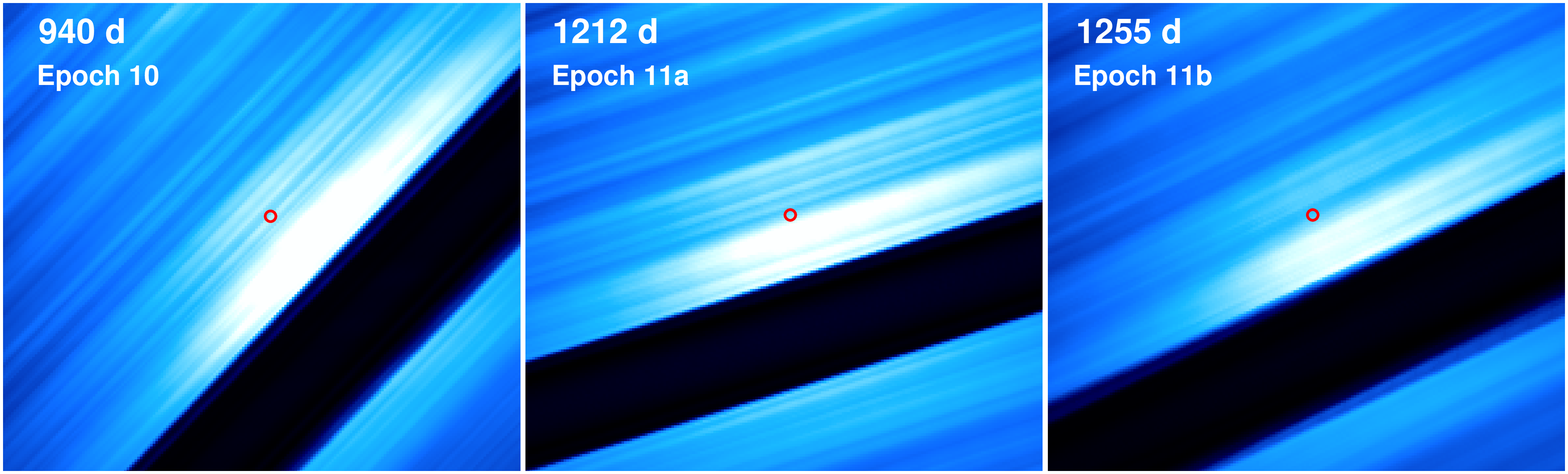}
\caption{Exposure maps for the last three sets of \textit{Chandra} observations, performed around 940~d (Epoch 10 in Table~\ref{tab1}), 
1212~d (Epoch 11a), and 1255~d (Epoch 11b) after the merger.
The position of GW170817 is marked by the red circle.}
\label{fig:expomap}
\end{center}
\end{figure*}

\begin{table*}
\centering
\caption{Log of \textit{Chandra} X-ray observations of GW170817.}
\label{tab3}
\begin{tabular}{cccccc}
\hline
\hline
ObsID & T-T$_0$  &  Exposure  & Count rate$^{a}$  & ECF$^{b}$ & Flux$^{c}$    \\
      &   (d)    &  (ks)      & [0.5-7.0 keV]     &           & [0.3-10 keV]         \\
\hline
18955  & 2.3  & 24.6  &  $<$2.5  & 1.65 &  $<$4.1   \\[0.9mm]
 19294  &  9.2     &   49.4    &   $2.9_{-0.7}^{+0.9}$ &   1.65  & $4.7^{+1.5}_{-1.2}$   \\[0.9mm]
 20728  &   15.4    &   46.7    & $3.8_{-0.9}^{+1.0}$ &   1.64    & $6.2^{+1.7}_{-1.5}$ \\[0.9mm]
 18988  &   15.9    &   46.7    &  $3.0_{-0.8}^{+1.0}$  &   1.65  &  $5.0^{+1.6}_{-1.2}$  \\[0.9mm]
 20860  &   108.0 &   74.1    & $15.0_{-1.5}^{+1.4}$  &   1.67    & $25^{+2}_{-2}$ \\[0.9mm]
 20861  &   111.1    &  24.7     & $15_{-2}^{+3}$ &  1.68    & $25^{+4}_{-4}$   \\[0.9mm]
 20936  &   153.6    &  31.7     & $20_{-3}^{+3}$ &  1.68    &  $33^{+4}_{-4}$  \\[0.9mm]
20938   &   157.1    &   15.9    &  $19_{-3}^{+4}$        & 1.68    & $32^{+6}_{-6}$   \\[0.9mm]
 20937  &   158.9    &  20.8     &  $15_{-3}^{+3}$        &  1.69  &   $25^{+5}_{-4}$ \\[0.9mm]
20939   &   159.9    &  22.2     & $11_{-2}^{+2}$         &  1.69   &  $18^{+4}_{-4}$  \\[0.9mm]
20945   &  163.7     &  14.2     &  $11_{-3}^{+3}$         &  1.69   &   $18^{+5}_{-4}$  \\[0.9mm]
21080   &   259.2    &   50.8    &  $7.8_{-1.2}^{+1.4}$         &  1.71   &   $13^{+2}_{-2}$  \\[0.9mm]
 21090  &  260.8     &  46.0     & $8.6_{-1.3}^{+1.5}$           &   1.71   & $15^{+3}_{-2}$   \\[0.9mm]
 21371  &   358.6    &   67.2    &  $5.2_{-0.9}^{+1.0}$          &   1.73   & $9.0^{+1.7}_{-1.5}$   \\[0.9mm]
21322   &    581.0   &   35.1    &  $1.5_{-0.6}^{+0.8}$         &   1.80   &  $2.7^{+1.5}_{-1.1}$  \\[0.9mm]
 22157  &    581.9   &    38.2   &   $2.0_{-0.7}^{+0.9}$         &   1.80   & $3.5^{+1.6}_{-1.2}$   \\[0.9mm]
 22158  &   583.6    &   24.9    &  $1.7_{-0.8}^{+1.1}$         &   1.80    & $3.1^{+1.9}_{-1.4}$ \\[0.9mm]
 21372  &  740.3     &   40.0    &  $0.5_{-0.3}^{+0.5}$         &   1.83    & $0.9^{+1.0}_{-0.6}$ \\[0.9mm]
22736   &   742.2    &   33.6    &  $1.3_{-0.6}^{+0.8}$         &   1.85    & $2.3^{+1.5}_{-1.0}$ \\[0.9mm]
 22737  &   743.1    &  25.2     & $2.1_{-0.9}^{+1.2}$          &  1.95    & $3.9^{+2.1}_{-1.6}$ \\[0.9mm]
 21323  &  935.5    &  24.3     &   $1.3_{-0.7}^{+1.0}$      &   1.92  &   $2.5^{+1.9}_{-1.2}$  \\[0.9mm]
 23183 &   939.1    &   16.3    &    $<$3.6      &   1.90   &  $<$6.8 \\[0.9mm]
 23184  &   940.6    &  19.8     &  $1.1_{-0.8}^{+1.2}$         &   1.91   & $2.1^{+1.5}_{-1.0}$  \\[0.9mm]
 23185  &    941.6   &    36.2   & $0.6_{-0.4}^{+0.6}$          &   1.91  &  $1.1^{+1.1}_{-0.7}$  \\[0.9mm]
22688  &    1209.7   &   29.7    &  $0.7_{-0.4}^{+0.7}$         &  1.99   &  $1.4^{+1.5}_{-0.9}$ \\[0.9mm]
24887   & 1212.3      &  26.7     &  $0.8_{-0.5}^{+0.8}$         &  1.99   &  $1.5^{+1.5}_{-0.9}$  \\[0.9mm]
  24888 &   1213.1    &   17.8    &  $2.4_{-1.1}^{+1.5}$         &  2.0    & $5^{+3}_{-2}$  \\[0.9mm]
 23889 &   1214.0    &  16.8     &   $1.9_{-1.0}^{+1.4}$        &   1.99   & $3.8^{+3}_{-1.9}$  \\[0.9mm]
 23870  &  1250.1     &   38.5    &  $1.1_{-0.5}^{+0.7}$         & 1.99    & $2.2^{+1.4}_{-1.0}$   \\[0.9mm]
 24923  &  1258.0     &   29.7    &  $<$2.0        &  1.99   & $<$3.9 \\[0.9mm]
  24934 &  1258.7     &  29.7     &  $<$2.6        &  2.0   & $<$5.3  \\[0.9mm]
\hline
\end{tabular}
\begin{flushleft}
     \quad \footnotesize{$^a$ Count rates are  in units of 10$^{-4}$ cts s$^{-1}$. All the values are corrected for PSF losses.Upper limits are 3\,$\sigma$.} \\
     \quad \footnotesize{$^b$ ECFs are in units of 10$^{-11}$ erg cm$^{-2}$ ct$^{-1}$} \\
     \quad \footnotesize{$^c$ Fluxes in units of 10$^{-15}$ erg cm$^{-2}$ s$^{-1}$. Values are corrected for Galactic extinction. Upper limits are 3\,$\sigma$.} \\
     
\end{flushleft}
\end{table*}

\begin{table*}
\centering
\caption{Flux densities from the latest radio observations of GW170817 by ATCA and VLA. The upper limits are at the $3\sigma$ level.}
\label{tab4}
\begin{tabular}{ccccccc}
\hline
\hline
MJD & T-$T_0$ & Telescope & Configuration & Project & Frequency & Flux density \\
(d) & (d) & & & & (GHz) &  ($\mu$Jy)   \\
\hline
59167 & 1185 &  ATCA & 6B &C3240 & 2.1 & $<51$  \\
59265 & 1283  & ATCA&  6D &C3240&  2.1 &$<75$ \\
59310 & 1328 & ATCA & 6D&C3240 &  2.1 & $<54$  \\
59312 & 1330 & ATCA & 6D &C3240&  5.5 & $<44$  \\
59312 & 1330 & ATCA & 6D &C3240& 9.0 & $<31$ \\
\hline
59198 & 1216 & VLA & A&SL0449 &3.0 & $<10.2$  \\
59210 & 1228 &  VLA& A &SL0449& 3.0 &$<9.9$  \\
59247 & 1265 & VLA &A &SM0329 &3.0 & $<8.7$  \\
59222$^{a}$ & 1240 & VLA &A & SL0449/SM0329 & 3.0& $<5.7$  \\
59255 & 1273  & VLA&A &SM0329 & 15.0 & $<5.3$  \\
\hline
\end{tabular}
\begin{flushleft}
     \quad \footnotesize{$^{a}$ This represents the exposure averaged MJD epoch from MJDs 59198,  59210, and
 59247.}
     
\end{flushleft}
\end{table*}
  
\bsp
\label{lastpage}
\end{document}